\documentclass[twocolumn]{article}

\usepackage[margin=1.8cm]{geometry}

\usepackage{caption}
\usepackage{subcaption}
\usepackage{graphicx}
\usepackage{dcolumn}
\usepackage{bm}
\usepackage{soul}
\usepackage{amsmath,amssymb,amsfonts,dsfont,xspace,relsize,mathtools,amsthm}
\usepackage[monochrome]{xcolor}
\usepackage{bbm}
\usepackage{physics}
\usepackage{ulem}
\usepackage{hyperref}
\hypersetup{
	colorlinks = true,
	urlcolor   = blue,
	linkcolor  = blue,
	citecolor  = red
}
\usepackage{algorithm}
\usepackage{algpseudocode}
\usepackage{float}
\usepackage{siunitx}
\usepackage[shortlabels]{enumitem}
\usepackage{tikz}

\newcommand{\ii}{{\rm i}}

\newcommand{\E}{\mathbb{E}}
\newcommand{\C}{\mathbb{C}}

\newcommand{\cA}{{\mathcal{A}}}

\newcommand{\cP}{{\mathcal{P}}}

\newcommand{\dketbra}[1]{|{#1}\rangle\langle{#1}|}

\newcommand{\ssection}[1]{\textit{#1} --}
\newcommand{\sign}[1]{{\rm sign}(#1)}
\newcommand{\mse}[1]{{\rm MSE}_{#1}}

\usepackage{authblk} \title{Private Remote Phase Estimation over a Lossy Quantum Channel} 

\author[1]{Farzad Kianvash} \author[2,3,4]{Marco Barbieri} \author[1]{Matteo Rosati} \affil[1]{Dipartimento di Ingegneria Civile, Informatica e delle Tecnologie Aeronautiche, Università degli Studi Roma Tre, Via della Vasca Navale 79, 00146 Rome, Italy} \affil[2]{Dipartimento di Scienze, Università degli Studi Roma Tre, Via della Vasca Navale 84, 00146 Rome, Italy} \affil[3]{Istituto Nazionale di Ottica - CNR (CNR-INO), Via Nello Carrara 1, Sesto F.no 50019, Italy} \affil[4]{INFN, Sezione di Roma Tre, Via della Vasca Navale 84, 00146 Rome, Italy}

\date{}

\begin{document}
\maketitle
\begin{abstract}
Private remote quantum sensing (PRQS) aims at estimating a parameter at a distant location by transmitting quantum states on an insecure quantum channel, limiting information leakage and disruption of the estimation itself from an adversary. Previous results highlighted that one can bound the estimation performance in terms of the observed noise. However, if no assumptions are placed on the channel model, such bounds are very loose and severely limit the estimation. We propose and analyse a PRQS using, for the first time to our knowledge, continuous-variable states in the single-user setting. Assuming a typical class of lossy attacks and employing tools from quantum communication, we calculate the true estimation error and privacy of our protocol, both in the asymptotic limit of many channel uses and in the finite-size regime. Our results show that a realistic channel-model assumption, which can be validated with measurement data, allows for a much tighter quantification of the estimation error and privacy for all practical purposes. 
\end{abstract}

\ssection{Introduction} Quantum sensing and cryptography are among the leading applications of modern quantum technologies. In recent years, with the rise of small quantum networks~\cite{Wehner2018,Rohde2025}, researchers have studied the potential of combining the unique features and advantages of both technologies into a single application. 
Enter private remote quantum sensing (PRQS), a technique that allows a quantum provider to estimate physical quantities at a distant location, connected via an insecure quantum communication channel~\cite{Huang2019a,Takeuchi2019,Yin2020,Shettell2022a,Moore2023,He2024,Ho2024,Hassani2024,Bugalho2024,Bizzarri2025}, with applications to biometry, remote medicine and, more generally, to the remote handling of sensitive data. In this client-server scenario, Alice owns quantum hardware that can produce probes for estimation, while Bob, the other legitimate user, sits at a distant location where the estimation has to take place. Thus, Alice seeks to carry out the estimation remotely, ensuring that Eve, who controls the quantum communication channel, gathers as little information as possible about the estimated parameter and, at the same time, does not spoil the legitimate estimation procedure.

\begin{figure}
    \centering
    \includegraphics[width=\columnwidth]{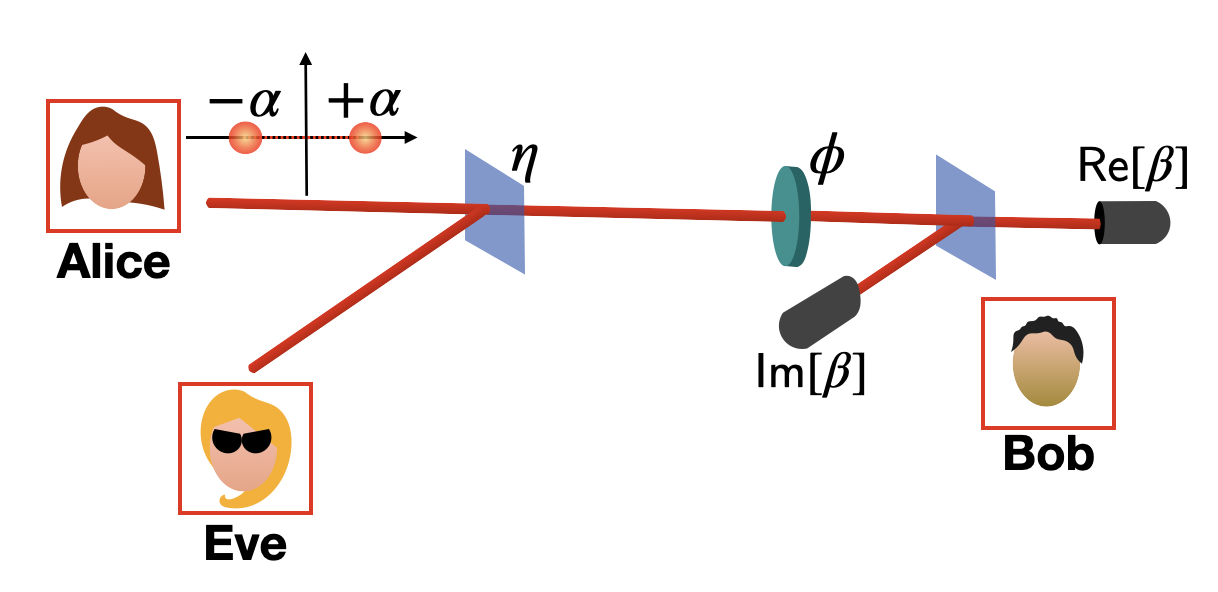}
    \caption{Schematic depiction of our PRQS protocol. Alice wants to estimate a phase $\phi$ at Bob's location, out of her reach. She thus delegates the measurement to Bob, who performs heterodyne on the transmitted probes. Eve compromises the security of the protocol by tapping part of the signal, in order to learn the initial phase, and can also access any classical communication between the parties.  }
    \label{fig:1}
\end{figure}

The PRQS problem has been well-studied using discrete-variable (DV) states in the single-user setting~\cite{Huang2019a,Takeuchi2019,Yin2020,Shettell2022a,Moore2023,He2024,Ho2024,Bizzarri2025}, employing tools from quantum cryptography for the analysis of metrologically relevant quantities such as the estimation bias and variance. In the related, but distinct setting of multi-user distributed sensing, research has focused instead on guaranteeing privacy with respect to other users rather than to an eavesdropper~\cite{Hassani2024,Bugalho2024,DeJong2025,Alushi2025,Junior2025}. These studies have highlighted the importance of establishing specific figures of merit for the task at hand, in order to put the notions of privacy and integrity on the quantitative level. However, this endeavour has revealed how difficulties may arise in bringing together concepts from cryptography and metrology. In particular, building on previous works in the single-user case, Ref.~\cite{Bizzarri2025} provided a threshold-based mechanism to check the probe's quality before proceeding to estimation and, with an experimental implementation, uncovered a major shortcoming of current PRQS approaches: while general limits can be inferred on the sensing performance by analyzing the transmitted probes, these are typically loose, as they have to account for any form of noise added to the probe by Eve. Pursuing an analogy with secure communications, insight in such cases are gained by inspecting specific attack models.

In this work we tackle exactly this issue, proving that the assumption of a specific noise model yields an accurate quantification of Eve's actions. While in theory this effectively restricts Eve's attack class, in practical cryptographic implementations it is common to work with reasonable assumptions on the channel model, which can then be regularly checked using the measurement data gathered for channel parameter estimation. 
Our work, to the best of our knowledge, pioneers PRQS protocols based on continuous-variable (CV) states in the single-user setting (for the distinct multi-user setting see instead the very recent Refs.~\cite{Alushi2025,Junior2025}). Similarly to the DV case where PRQS schemes are based on seminal QKD protocols such as BB84 and E91~\cite{Bennet1984,Ekert1991}, our proposed CV protocol is reminiscent of GG02~\cite{Grosshans2002}.
However, in order to provide a realistic protocol and avoid over-simplifications connected with a continuous signal distribution, at variance with GG02 we consider a discrete constellation of probe states~\cite{Ghorai2019,Lin2019a,Denys2021,Kanitschar2023}. As a consequence, tackling Eve's attacks requires yet again the use of tools from quantum communication: we originally integrate results from minimum-error state discrimination~\cite{Rosati17c,Bilkis2020a,Bilkis2021a,Rosati2023,Rosati2025,Kianvash2025} to quantify the attacker's estimation performance under a discrete modulation of the original probe states.
Finally, our protocol analysis is valid for any number of channel uses, enabling a complete understanding of security and estimation performance both in the asymptotic limit and in the finite-size regime, which is notoriously hard to tackle in CV-QKD. Therefore, our results are a valuable asset for informing future deployments in real networks.

\ssection{The protocol} We consider the setting depicted in Fig.~\ref{fig:1}. At each round of the protocol, Alice transmits a single-mode coherent state of the electromagnetic field, $\ket\alpha_i=e^{ a^\dagger \alpha_i- a \alpha_i^*}\ket0$,  over an insecure quantum channel to Bob, where $ a,a^\dagger$ are the mode photon-annihilation and -creation operators satisfying the canonical commutation relations $[ a, a^\dagger]=1$. In order to enforce security, Alice picks the coherent-state complex amplitude from a discrete set with fixed mean photon number $\alpha^2$, i.e., $\alpha_i\in\cA=\{\alpha e^{\ii \theta_k}:\alpha>0\}_{k=1}^{A}$. Bob encodes an unknown phase $\phi$ on the received states via the unitary $ R(\phi) = e^{-i\phi  a^\dagger  a}$ and then performs heterodyne measurements, represented by the positive-operator-valued-measurement $\{\frac1\pi \dketbra{\beta}\}_{\beta\in\C}$, obtaining  in the $i$-th run the outcome $\beta_i$. While the scenario in itself is quite general, we assume that Eve can perform only passive attacks:  the noise model is a bosonic pure-loss channel, mapping $\ket{\alpha_i}_A\mapsto\ket{ \alpha_i \sqrt\eta}_B$; therefore, the remaining part of the probe state is intercepted by Eve, i.e., $\ket{\alpha_i \sqrt{1-\eta}}_E$. This restriction has the advantage of providing a clear setting to bound the performance of the PRQS protocol, while retaining a certain level of practical interest.
After $N$ rounds, Bob first enters a {\footnotesize CHECK} phase, employing the measured data to evaluate which fraction of the signal was intercepted by Eve, and determine if that guarantees sufficient privacy upon disclosing the measurement outcomes. If the {\footnotesize CHECK} is passed, then Bob publicly announces his outcomes $\{\beta_i\}_{i=1}^N$, allowing Alice to proceed with the {\footnotesize ESTIMATION} phase. 

We stress that, since our protocol does not rely directly on the security of GG02, it enables us to privilege considerations on practicality, pointing to the use of a more controlled discrete modulation of Alice's phase~\cite{Bilkis2020a,Notarnicola22,Kianvash2025}. While this compromises the Gaussianity of the overall state, hence requiring greater care in secret key analysis~\cite{Denys2021,Kanitschar2023}, we will show how our PRQS protocol is immune from such limitations, thanks also to the passive-attack assumption.

\ssection{Binary-phase alphabet}
In order to illustrate what happens in each phase specifically, we assume for simplicity the case of a symmetric binary-phase-shift-keyed (BPSK) alphabet $\cA=\{\pm \alpha: \alpha\geq0\}$. 
In the {\footnotesize ESTIMATION} phase, Alice corrects Bob's data with her knowledge of the initial phases, obtaining the dataset $\{\beta_i^A = \beta_i \cdot s_i\}_{i=1}^N$, where $s_i=\sign{\alpha_i}$; then, she estimates the phase via the estimator $\hat\phi^A = \arg\frac{1}{N}\sum_{i=1}^{N} \beta^A_i$, which is maximum-likelihood, as shown in the Supplementary Material (SM).
Alice's performance is quantified by the mean-square error (MSE). In the local-estimation setting, where the true value of $\phi$ is approximately known,  one studies the local MSE, averaging with respect to the outcome distribution, i.e., $ \mse{A}(\phi) = \E_{\{\beta_i^A\}|\phi}{(\hat\phi^A-\phi)^2}$;
instead, in the Bayesian setting one studies the average MSE with respect to the phase distribution, i.e., $    \mse{A} =  \E_{\phi}\mse{A}(\phi)$.
In both cases, the MSE is a function of the received coherent-state amplitude $\sqrt{\eta}\,\alpha$, a priori unknown to the legitimate users due to the loss introduced by Eve.
Clearly, in trying to estimate $\phi$, Eve will try a similar procedure but will first need to identify the phase of $\alpha_i$ chosen by Alice, via a BPSK discrimination experiment~\cite{Bilkis2020a}. Eve's error is then quantified by the Helstrom bound~\cite{helstromBOOK,Bilkis2020a}: $P_e^{\mathrm{Hel}}=\tfrac12\!\left(1-\sqrt{1-e^{-4(1-\eta)\alpha^2}}\right)$, i.e., with probability $p(\eta)=P_e^{\mathrm{Hel}}$ Eve will identify the wrong phase $\hat s_i = -\sign{\alpha_i}$, while with probability $q(\eta)=1-p(\eta)$ she will identify the correct one $\hat s_i=\sign{\alpha_i}$. Eve's corrected outcomes are $\{\beta_i^E = \beta_i \cdot \hat s_i\}_{i=1}^N$,  and her estimator $\hat\phi^E$ and $\mse{E}$ are analogous to those of Alice.

Therefore, in the {\footnotesize CHECK} phase Bob needs to ensure that Eve's estimation will be more noisy than Alice's. The degree to which this is true is defined as privacy:
\begin{equation}
    \label{eq:privacy_def}
    \cP = 1-\frac{\mse{A}}{\mse{E}}.
\end{equation}
Observe that $\cP\in[0,1]$ if $\mse{E}\geq\mse{A}$, and it attains its maximum when $\mse{A}/\mse{E}\rightarrow 0$, i.e., when Alice's error is vanishingly small with respect to that of Eve. 
In light of these reasons, given the measurement data $\{\beta_i\}_{i=1}^N$, Bob estimates the privacy \eqref{eq:privacy_def} with the maximum-likelihood estimator $ \hat \cP = \cP|_{\eta=\abs{\frac{1}{N\alpha}\sum_{i=1}^{N} \beta^A_i}^2}$ and checks whether it respects a predefined threshold, i.e., $\hat\cP\geq 1-\epsilon$, where $\epsilon$ was previously agreed upon by Alice and Bob, also taking into account the uncertainties on the estimated privacy . If this holds true, then the check passes, otherwise the protocol is aborted, in line with DV protocols~\cite{Huang2019a,Shettell2022a,Bizzarri2025}. Note that, in the latter case, Eve will not be able to estimate the phase at all, since she does not have access to the measurement data. 

\ssection{MSE analysis} Bob's heterodyne outcome in round $i$ can be written as $ \beta_i = \sqrt{\eta}\, s_i\, \alpha + n_i$ where $n_i \sim \mathcal{CN}(0,1)$ iid, and $\mathcal{CN}(m,\sigma^2)$ denotes a circularly symmetric complex Gaussian random variable 
with mean $m$ and variance $\sigma^2$. 
Alice forms the sign-aligned variables $    \beta^{A}_{i}=r(\phi)+\tilde n_i$, where $r(\phi)= \sqrt{\eta}\,\alpha e^{\ii\phi}>0$ and $\tilde n_i\equiv s_i n_i\sim\mathcal{CN}(0,1),$ then she calculates the signed average $\bar\beta^A =\frac1N \sum_{i=1}^N \beta_i^A\sim\mathcal{CN}\!\Big(r(\phi),\frac{1}{N}\Big)$.
Therefore, Alice's local MSE can be written as
\begin{equation}\label{eq:alice_mse_local}
  \mse{A}(\phi)=\int_\C d^2 z\, (\theta-\phi)^2 p_{\bar{\beta}^{A}}(z|\phi),
\end{equation}
where $p_{\bar{\beta}^{A}}(z|\phi)=\frac{N}{\pi}\exp\!\big(-N\abs{z-r(\phi)}^{2}\big)$ is the probability density function of the signed average and $\theta=\arg z$ is the value of the estimator $\hat\phi^A$ when $\bar\beta^A=z$. Now, let us note that $\mse{A}(\phi)$ depends only on the difference $\theta-\phi$, up to an irrelevant global-phase change in the exponent of $p_{\bar{\beta}^{A}}(z|\phi)$; as a consequence, the local MSE is independent of the phase $\phi$ to be estimated and it equals the average MSE. We conclude that the estimator's performance is the same both in the local and Bayesian setting, independently of the prior, 
and is given by \eqref{eq:alice_mse_local} with $\phi=0$. Finally, by setting $z=\rho e^{\ii \theta}$ and carrying out the radial integral we obtain $\mse{A} = \int_{-\pi}^\pi d\theta \,\theta^2 p_{\hat \phi}(\theta;r)$, with
\par \small \begin{align}
p_{\hat\phi}(\theta;r)
&=\frac1{2\pi}e^{-N r^2} \label{eq:alice-angle-pdf-closed}\\
&\cdot\left[1+\sqrt{\pi N} r\cos\theta\;
e^{N (r\cos\theta)^2}\,
\left(1+\operatorname{Erf}(\sqrt{N} r\cos\theta)\right)\right]\nonumber
\end{align}
\normalsize the posterior probability density of Alice's estimator, $r=r(0)$, and Erf(x) the error function, as shown in the SM. Interestingly, note that, since \eqref{eq:alice-angle-pdf-closed} is an even function of $\theta$, the estimator is unbiased for any $N$. 

Similarly, Eve's sign-aligned variables can be written as $ \beta^E_i=\hat{s}_i\beta_i=D_i r(\phi)+\tilde n_i$, where $ \tilde n_i\sim\mathcal{CN}(0,1)$,
$D_i=\hat{s}_i s_i\in\{\pm1\}$ and $\Pr(D_i=-1)=p(\eta)$; hence their probability density is a mixture of Gaussians weighted by the Helstrom probabilities $p(\eta)$, $q(\eta)$. As shown in the SM via the characteristic function, the probability density of Eve's signed average, $\bar{\beta}^{E}\equiv \frac{1}{N}\sum_{i=1}^{N} \beta^E_i$, is also a mixture of Gaussians with binomial weights:
\par \small \begin{equation}
p_{\bar{\beta}_E}(z|\phi)= \E_{k\sim{\rm Bin}(N,p(\eta))}\exp\!\Big(-N\,\left|z-r(\phi)\Big(1-\frac{2k}{N}\Big)\right|^{2}\Big),
\end{equation}
\normalsize where ${\rm Bin}(N,p(\eta))$ is the binomial distribution of the number of errors $k$ in $N$ discrimination experiments with single-error probability $p(\eta)$. We conclude that, as in Alice's case, Eve's local MSE is $\phi$-independent and therefore equal to the average MSE:
\begin{equation}
\begin{aligned}
    \mse{E} 
    &= \E_{k\sim{\rm Bin}(N,p(\eta))} \mse{A}\big|_{r\mapsto r_k}.\label{eq:eve_mse_final}
    \end{aligned}
\end{equation}
Interestingly, \eqref{eq:eve_mse_final} shows that Eve's MSE can be expressed as a binomial expectation of the MSE that Alice would have if she started with a different coherent-state amplitude $r_k=\sqrt\eta\, \alpha(1-\frac{2k}{N})$; indeed, the latter corresponds to Eve's observed average amplitude over the $N$ rounds, when she commits exactly $k$ discrimination errors. 

\begin{figure*}[!t]
    \centering
    \includegraphics[width=.8\linewidth]{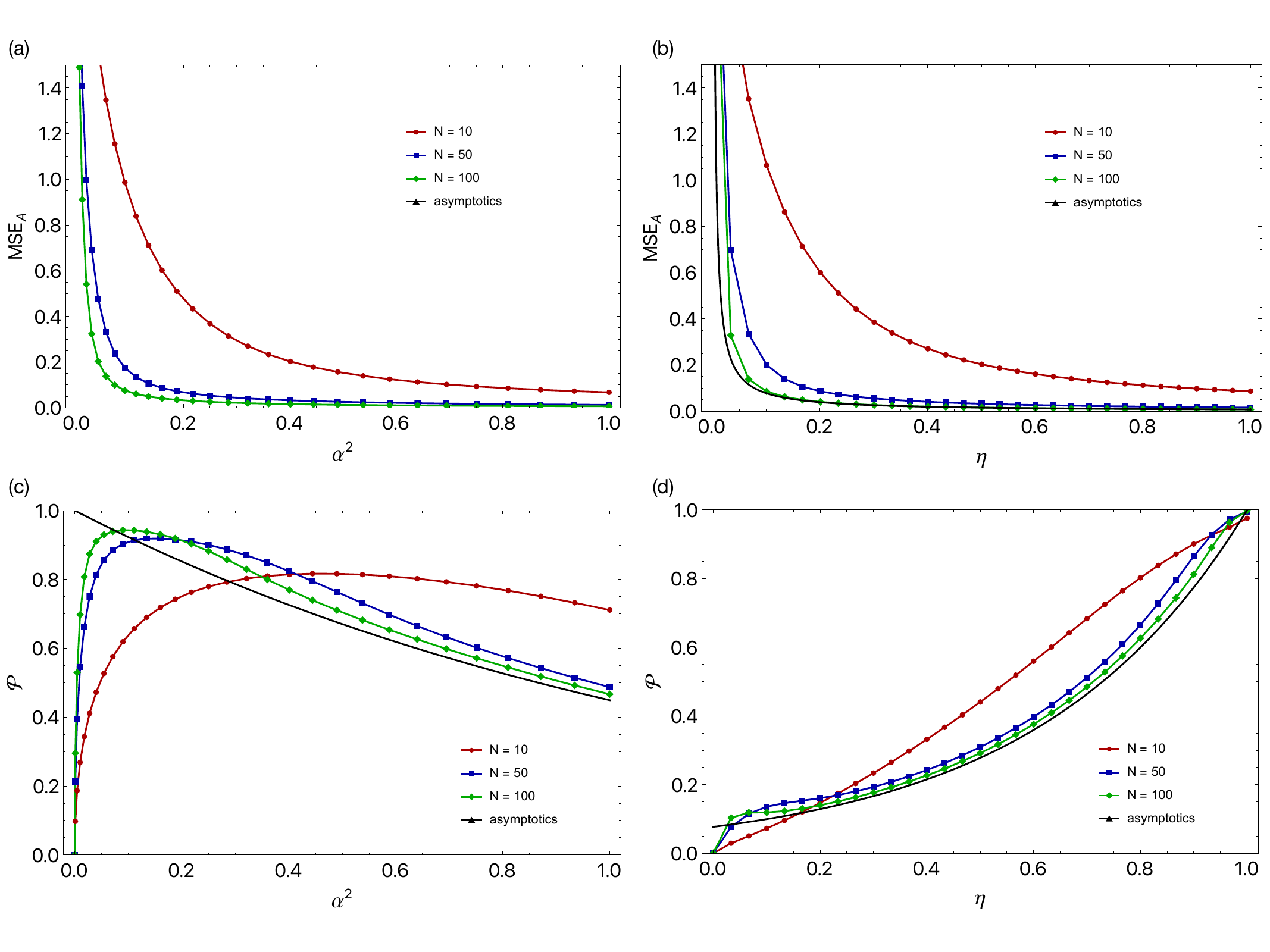}
    \caption{Plots of the finite-size and first-order asymptotic ($N=100$) behaviour of the estimation error $\mse{A}$ (a,b) and privacy $\cP$ (c,d), as a function of the initial probes' mean-photon number $\alpha^2$ (a,c) and channel transmissivity $\eta$.}
    \label{fig:2}
\end{figure*}

\ssection{Asymptotic performance} In the large-$N$ limit, we can obtain analytical expressions for the MSE's and the privacy parameter by linearizing the estimator and approximating the binomial distribution with a Gaussian. Indeed, Eve's MSE can be written as an expectation over three random variables, obtained via averaging of Eve's data: $\mse{E} = \E\arg^2(r \bar D + X + \ii Y)$, where $\bar D = \frac1N\sum_{i=1}^N D_i$ is binomial-distributed, while $X$ and $Y$ are the real and imaginary parts of the complex-normal-distributed random variable $\frac1N \sum_{i=1}^N \tilde n_i$. Expanding the $\arg$ function to the second order, we obtain
\par \small\begin{equation}
\mse{E}
=\mathbb{E}\!\left[\left(\frac{Y}{r\bar D + X}\right)^2-
\frac{2}{3}\left(\frac{Y}{r\bar D + X}\right)^{\!4}\right]+ O(N^{-3}).
\label{eq:mseE-core-corrected}
\end{equation}
\normalsize Since the three variables are independent, the expectation with respect to $Y$ can be carried out straightforwardly. Instead, the expectation of the quantities in the denominator can be approximated via a Taylor series of the probability distribution, also known as Edgeworth expansion~\cite{Kolassa2006}, obtaining (see SM):
\par \small \begin{equation}
\mathrm{MSE}_E
\sim\frac{1}{2N\,r^{2}(1-2p(\eta))^{2}}\left(1+\frac{1+24r^{2}p(\eta)(1-p(\eta))}
     {4 N\,r^{2}(1-2p(\eta))^{2}}\right)
\label{eq:mseE-final-rp}
\end{equation}
\normalsize up to $O(N^{-3})$, which characterizes the precision beyond the usual asymptotic assessment based on the Cram\'er-Rao bound. Note that Alice's MSE can be recovered in the same limit by setting $p(\eta)=0$ above, i.e., 
\begin{equation}\label{eq:alice_mse_asy}
    \mse{A}\sim\frac{1}{2N\,r^{2}}\left(1+\frac{1}
     {4N\,r^{2}}\right).
\end{equation}
Therefore, the privacy behaves asymptotically as
\begin{equation}
    \cP \sim 1- (1-2p(\eta))^2 \left(1-\frac{(6r^{2}+1)p(\eta)(1-p(\eta))}
     {N\,r^{2}(1-2p(\eta))^{2}}\right)
\end{equation}
up to $O(N^{-2})$. In particular, as $N\rightarrow\infty$ the privacy tends to $\cP_{\infty} = \exp(-4 \alpha^2 (1 - \eta))$ for finite $\alpha$. Furthermore, since we now have an analytical expression $\cP_\infty(\eta)$, a monotonic function of $\eta$, in the large-$N$ limit a threshold on the privacy can be established in terms of a confidence interval. Namely, Alice and Bob can easily choose a threshold $\epsilon$ and a confidence interval $\delta$ such that ${\rm Pr}(\cP_\infty(\eta)\geq 1-\epsilon|\{\beta_i\}_{i=1}^N)\equiv {\rm Pr}(\eta\geq \eta(\epsilon)|\{\beta_i\}_{i=1}^N)\geq 1-\delta$.

\normalsize  We conclude the asymptotic analysis by observing that the minimum MSE attainable with coherent-state probes scales, for large $N$, as $\mse{A}/2$, and the optimal measurement is homodyne~\cite{helstromBOOK}, aligned orthogonally to the initial coherent-state phase. Note however that Alice and Bob need to estimate also the loss $\eta$, hence the estimation performance has to be evaluated in a multi-parameter setting, which does not allow for optimising the uncertainties on individual parameters. Nevertheless, previous works point to the optimality of coherent-states and heterodyne measurements for the joint phase- and amplitude- estimation with single-mode probes ~\cite{Genoni2013}. Furthermore, the use of non-binary constellations will immediately favour heterodyne over homodyne. 

\ssection{Finite-size behaviour} The previously obtained MSE and privacy expressions can be studied numerically for any finite $N$, providing a full characterization of the protocol's performance in the finite-size regime. In Fig.~\ref{fig:2} we plot Alice's MSE and the privacy as a function of the probe's mean-photon number and the channel transmissivity. We observe that $\mse{A}$ decreases both with $\alpha^2$ and $\eta$; similarly, $\cP$ increases with $\eta$. This can be understood since in all three cases the figure of merit is enhanced by a larger received probe mean-photon number. On the other hand, it appears that, for any finite $N$, there exists an optimum value of $\alpha^2$ that maximizes the estimation privacy. This is due to the trade-off between decreasing $\mse{A}$, which requires large $\alpha^2$, and decreasing $\mse{A}/\mse{E}$, which requires small $\alpha^2$. Below a certain threshold number of photons, the higher noise experienced by Eve cannot compensate for the decrease in estimation error experienced by Alice. Furthermore, we note that for large photon number the asymptotic expression is already valid at moderate $N>50$. However, the approximation breaks down for sufficiently small energy, where a more careful expansion at fixed $N \alpha^2$ should be undertaken to analyze the small-$\alpha$ behaviour, using similar methods as above.

\ssection{Conclusions} We have introduced and fully characterized the first single-user CV-RPQS protocol under the assumption of passive photon-splitting attacks. This allowed us to derive analytical results with the use of a discrete-modulation, usually hard to tackle, by bounding the attacker's discrimination capabilities. We studied both the asymptotic and finite-size regime, introducing the concept of estimation privacy and studying its connection with the estimation error. Our results show that, for finite number of rounds, there is an optimal probe mean-photon number that ensures the highest privacy while maintaining good estimation performance, similarly to the optimal modulation of GG02. Our work illustrates the merits of inspecting technologically relevant specific attacks for more stringent bounds in PRQS.

\ssection{Acknowledgements} We thank Gabriele Bizzarri for useful discussions on PQRS. 
M.R. acknowledges support from the project PNRR -Id MSCA 0000011-SQUID-CUP F83C22002390007 (Young Researchers) - Finanziato dall'UE - NextGeneration EU. 
This work has been realised with the support of  the PRIN project PRIN22-RISQUE-2022T25TR3 of the Italian Ministry of University. 
M.B. acknowledges support by Rome Technopole Innovation Ecosystem (PNRR grant M4-C2-Inv) and  MUR Dipartimento di Eccellenza 2023-2027.

\onecolumn
\appendix

\section{Full derivation of the MSE for BPSK alphabet}\label{sec:alice-binary}
We consider the protocol described in the main text with a BPSK alphabet $\{\pm\alpha:\alpha>0\}$. As we have already shown that the MSE is $\phi$-invariant, without loss of generality we align the axis so that the unknown phase is $\phi=0$, and let $\alpha_i = s_i \alpha$ with $s_i \in \{+1,-1\}$. 
Bob's heterodyne outcome in round $i$ is modeled as
\begin{equation}
\beta_i = \sqrt{\eta}\, s_i\, \alpha + n_i, 
    \qquad n_i \sim \mathcal{CN}(0,1)\ \text{i.i.d.},
\end{equation}
where $\mathcal{CN}(m,\sigma^2)$ denotes a circularly symmetric complex Gaussian random variable 
with mean $m$ and variance $\sigma^2$, i.e.
\begin{equation}
    p_{Z}(z) = \frac{1}{\pi \sigma^{2}} 
    \exp\!\left(-\frac{|z - m|^{2}}{\sigma^{2}}\right), 
    \qquad z \in \mathbb{C}.
\end{equation}
In this notation, $\Re(Z)$ and $\Im(Z)$ are independent real Gaussian variables, each with mean 
$\Re(m)$ and $\Im(m)$ respectively, and variance $\sigma^{2}/2$.

\subsection{Alice's MSE}
Alice forms the sign-aligned variables
\begin{equation}
\beta^{A}_{i}\equiv s_i\beta_i=r+\tilde n_i,\qquad r\equiv \sqrt{\eta}\,\alpha>0,\quad \tilde n_i\equiv s_i n_i\sim\mathcal{CN}(0,1).
\end{equation}
The signed average is
\begin{equation}
    \bar{\beta}^{A}\equiv \frac{1}{N}\sum_{i=1}^{N} \beta^A_i\ \sim\ \mathcal{CN}\!\Big(r,\frac{1}{N}\Big).
\end{equation}
then Alice's estimator for the phase is $\hat{\phi}_A=\arg(\bar{\beta}^A)$. Next, we compute the mean square error (MSE) of the estimator.

Since $\bar{\beta}^{A}\sim\mathcal{CN}(r,1/N)$ with $r=\sqrt{\eta}\,\alpha>0$, its pdf is
\begin{equation}\label{eq:alice_prob_distro_z}
    p_{\bar{\beta}^{A}}(z)=\frac{N}{\pi}\exp\!\big(-N\abs{z-r}^{2}\big),\qquad z\in\mathbb{C}.
\end{equation}
Write $z=\rho e^{i\theta}$ with $\rho\ge 0$, $\theta\in(-\pi,\pi]$ and Jacobian
$\mathrm{d}x\,\mathrm{d}y=\rho\,\mathrm{d}\rho\,\mathrm{d}\theta$. Then
\begin{equation}
p_{\hat\phi_A}(\theta)
=\int_{0}^{\infty} p_{\bar{\beta}^{A}}(\rho e^{i\theta})\,\rho\,\mathrm{d}\rho
=\int_{0}^{\infty}\frac{N}{\pi}\exp\!\big(-N(\rho^{2}+r^{2}-2r\rho\cos\theta)\big)\,\rho\,\mathrm{d}\rho,
\label{eq:alice-angle-pdf-integral}
\end{equation}
and the mean-squared error of the phase estimator $\hat\phi_A=\arg(\bar{\beta}^{A})$ is
\begin{equation}
\mathrm{MSE}_{A}
=\int_{-\pi}^{\pi}\theta^{2}\,p_{\hat\phi_A}(\theta)\,\mathrm{d}\theta.
\label{eq:alice-mse-angle}
\end{equation}

The radial integral in \eqref{eq:alice-angle-pdf-integral} can be evaluated in closed form by completing the square. 
Let $\kappa\equiv N r^{2}=N\eta\alpha^{2}$, then we have
\begin{align}
    p_{\hat\phi_A}(\theta) & = \frac1\pi e^{-\kappa}\int_0^\infty d\rho\,\rho \exp(-\rho^2+2\sqrt{\kappa}\rho\cos\theta) = \frac1\pi e^{-\kappa\sin^2\theta}\int_0^\infty d\rho\,\rho \exp(-(\rho-\sqrt\kappa \cos\theta)^2) \\
    &= \frac1\pi \int_{-\sqrt{\kappa}\cos\theta}^{\infty}\! e^{-t^2}\Big(t+\sqrt{\kappa}\,\cos\theta\Big)\,\mathrm{d}t  = \frac{1}{2\pi} e^{-\kappa\sin^2\theta}\left(e^{-\kappa\cos^2\theta}+\sqrt{\pi \kappa} \cos\theta (1+{\rm Erf}(\sqrt\kappa \cos\theta))\right), 
    \end{align}
    which matches the expression of $p_{\hat\phi}(\theta;r)$ in the main text. 

\subsection{Eve's MSE}
\label{sec:eve-cf-mse}
Eve performs a Helstrom test on her environmental mode. Let
\[
P_e^{\mathrm{Hel}}=\tfrac12\!\left(1-\sqrt{1-e^{-4(1-\eta)\alpha^2}}\right),\qquad
r=\sqrt{\eta}\,\alpha>0,\qquad p=P_e^{\mathrm{Hel}},\ \ q=1-p.
\]
where we omit the dependence of the probabilities on $\eta$. With Eve's decision $\hat{s}_i\in\{\pm1\}$ and $D_i=\hat{s}_i s_i\in\{\pm1\}$, the corrected public data is
\begin{equation}
\beta^E_i=\hat{s}_i\beta_i=D_i r+\tilde n_i,\qquad
    \Pr(D_i=+1)=q,\ \Pr(D_i=-1)=p,\quad \tilde n_i\sim\mathcal{CN}(0,1),
\end{equation}
so 
\begin{equation}
    p_{\beta^E_i}(z)=q\,\frac{1}{\pi}e^{-|z-r|^2}+p\,\frac{1}{\pi}e^{-|z+r|^2}.
\end{equation}
Similar to Alice's case, Eve takes the sign-corrected average of Bob's public data 
\begin{equation}
    \bar{\beta}^{E}\equiv \frac{1}{N}\sum_{i=1}^{N} \beta^E_i,
\end{equation}
and her estimator is $\hat{\phi}_E=\arg(\bar{\beta}^E)$. Therefore, We are interested in the probability distribution of $\bar{\beta}^{E}$. For this, we make use of the characteristic function of a two-variable probability distribution defined as $\phi_Z(u_x,u_y)=\mathbb{E}[e^{i(u_x X+u_y Y)}]$, where $Z=X+iY$ is our complex random variable. Therefore, in our case
\begin{equation}
\phi_{\beta^E_i}(u_x,u_y)=e^{-\frac{1}{4}(u_x^2+u_y^2)}\Big[q\,e^{i r u_x}+p\,e^{-i r u_x}\Big].
\end{equation}
As $\bar{\beta}^E$ is the average of $N$ iid variables, using standard properties of the characteristic function we can write 
\begin{equation}
\phi_{\bar{\beta}_E}(u_x,u_y)=\Big(\phi_{\beta_i}\!\big(\tfrac{u_x}{N},\tfrac{u_y}{N}\big)\Big)^{\!N}
=\exp\!\Big(-\frac{u_x^2+u_y^2}{4N}\Big)\Big[q\,e^{i\frac{r}{N}u_x}+p\,e^{-i\frac{r}{N}u_x}\Big]^{\!N}.
\end{equation}
Expanding the power shows that $\bar{\beta}_E$ is a finite Gaussian mixture, with binomial coefficients:
\begin{equation}
p_{\bar{\beta}_E}(z)=\sum_{k=0}^{N}\binom{N}{k}q^{N-k}p^{k}\,
\frac{N}{\pi}\exp\!\Big(-N\,|z-r_k|^{2}\Big),
\qquad r_k=r\Big(1-\frac{2k}{N}\Big),
\end{equation}
which can be written as an average of Alice's probability distribution \eqref{eq:alice_prob_distro_z} with different $k$-dependent mean:
\begin{equation}
    p_{\bar\beta_E}(z) = \sum_{k=0}^N {\rm Bin}(k|N,p)\, p_{\bar{\beta}^{A}}(z)\Big|_{r\mapsto r_k}.
\end{equation}
By linearity, from the previous calculation in Alice's case it then follows the result of the main text.

\subsection{Optimality properties of the estimators}
It is straightforward to show that Alice's chosen estimator is maximum-likelihood. Given the sign-aligned variables $\{\beta_i^A\}_{i=1}^N$ we can write their joint probability density as
\begin{equation}
    p_{\{\beta_i^A\}_{i=1}^N}(z_1,\cdots,z_N) = \frac{1}{(\pi\sigma)^N} \exp\left(-\frac{1}{\sigma^2}\sum_{i=1}^N |z_i - r e^{\ii\phi}|^2\right)
\end{equation}
and we can write the log-likelihood function (for flat prior) as
\begin{equation}
    \ell(\phi|z_1,\cdots,z_N) \propto - \sum_{i=1}^N (\abs{z_i}^2 + r^2 -2r \Re{z_i e^{-\ii\phi}}) = {\rm const.} + 2r \abs{\sum_{i}z_i}\cos(\phi-\arg{\sum_i z_i}) . 
\end{equation}
Setting its derivative with respect to $\phi$ equal to zero we obtain the value of our estimator:
\begin{equation}
  \hat \phi_{\rm ml}= \arg\sum_{i=1}^N z_i \equiv \arg\frac1N\sum_{i=1}^N z_i.
\end{equation}
As a by-product, we also obtain the joint maximum-likelihood estimator for $r$, hence for the transmissivity:
\begin{equation}
    \hat \eta = \abs{\frac1{N\alpha}\sum_{i=1}^N z_i}^2.
\end{equation}
Finally, since $\cP(\eta)$ is a deterministic function, a maximum-likelihood estimator for $\cP$ is exactly $\hat\cP=\cP(\hat\eta)$.

\section{Large-$N$ limit}
We recall that Eve forms the corrected average
\begin{equation}
\bar{\beta}^{E} \;=\; \frac{1}{N}\sum_{i=1}^{N} \beta_i^{E}
\;=\; \frac{1}{N}\sum_{i=1}^{N}\!\big(D_i\,r+\tilde n_i\big)
\;=\; r\,\bar D \;+\; \bar n,
\qquad r\equiv \sqrt{\eta}\,\alpha>0.
\label{eq:eve-average}
\end{equation}
Hence $\bar n\sim\mathcal{CN}(0,1/N)$ and
\begin{equation}
\bar D \;\equiv\; \frac{1}{N}\sum_{i=1}^{N} D_i, 
\qquad 
\mathbb{E}[D_i]=1-2p,\quad \mathrm{Var}(D_i)=4p(1-p).
\end{equation}
Eve'ss estimator is $\hat\phi_E=\arg(\bar{\beta}^{E})$, and we wish to compute
\begin{equation}
\mathrm{MSE}_E\;\equiv\; \mathbb{E}\!\big[\hat\phi_E^{\,2}\big]
\end{equation}
for large $N$.

\subsection{Small-angle linearization at large $N$.}
Write $\bar n = X+iY$ with $X,Y\sim\mathcal{N}(0,\tfrac{1}{2N})$ independent, and note that
$\bar{\beta}^{E}=r\bar D + X + iY$. Then the estimator can be expressed as
\begin{equation}
\hat\phi_E \;=\; \arg(r\bar D + X + iY)
\;=\; \arctan\!\frac{Y}{r\bar D + X}.
\label{eq:exact-angle}
\end{equation}
For large $N$, $X,Y=O(N^{-1/2})$, while $\bar D = O(1)$, hence we may expand the arctangent around zero as
\begin{equation}
\hat\phi_E \;=\; \frac{Y}{r\bar D + X} \;-\; \frac{1}{3}\!\left(\frac{Y}{r\bar D + X}\right)^{\!3} \;+\; O\left(\left(\frac{Y}{r\bar D + X}\right)^{\!5}\right).
\label{eq:small-angle-corrected}
\end{equation}

Now note that the three random variables are independent and, in particular, it holds $\mathbb{E}[Y]=0$, $\E[Y^2]=1/(2N)$ and $\mathbb{E}[Y^4]=3/(4N^2)$, hence
\begin{equation}
\mathrm{MSE}_E
\;=\; \mathbb{E}\!\big[\hat\phi_E^{\,2}\big]
=
\frac{1}{2N}\,
\mathbb{E}\!\left[\frac{1}{(r\bar D + X)^{2}}\right]-
\frac{1}{2 N^2}\,\mathbb{E}\!\left[\frac{1}{(r\bar D + X)^{4}}\right]
\;+\; O(N^{-3}).
\label{eq:mseE-core-corrected}
\end{equation}

\subsection{Edgeworth (delta) expansion}
Let $\{W_i\}_{i=1}^N$ be i.i.d.\ random variables with mean
$\mu=\mathbb{E}[W_i]$, variance $\sigma^2=\mathrm{Var}(W_i)$, and higher cumulants
$\kappa_3,\kappa_4,$ etc. For the sample mean
\[
\bar W \;=\; \frac{1}{N}\sum_{i=1}^{N} W_i,
\]
we have the cumulant scalings
\[
\kappa_1(\bar W)=\mu,\qquad
\kappa_2(\bar W)=\frac{\sigma^2}{N},
\]
Then for any sufficiently smooth test function $g$, the Edgeworth (or delta-method) expansion of the expectation of $g(\bar W)$ around $\mu$ is~\cite{Kolassa2006}
\begin{equation}
\mathbb{E}\!\big[g(\bar W)\big]
= g(\mu)
+ \frac{1}{2}\,g''(\mu)\,\frac{\sigma^2}{N}
+ O\!\left(N^{-2}\right).
\label{eq:edgeworth-general}
\end{equation}
Here, the first term is the mean value $g(\mu)$, while the second one captures Gaussian fluctuations, and all higher cumulant contributions are contained in the remainder $O(N^{-2})$.

We now let
\begin{equation}
\bar W \;\equiv\; r\,\bar D + X \;=\; \frac{1}{N}\sum_{i=1}^{N} W_i,
\qquad W_i \;\coloneqq\; rD_i + X_i,
\end{equation}
where $X_i\sim\mathcal{N}(0,\tfrac{1}{2})$ are iid and independent of $D_i$.
Then, the random variables $W_i$ have the following cumulants:
\begin{align}
\mu &\;\equiv\; \mathbb{E}[W_i] \;=\; r(1-2p),\\[3pt]
\sigma^2 &\;\equiv\; \mathrm{Var}(W_i) \;=\; 4r^2p(1-p) + \tfrac{1}{2},
\end{align}
while higher cumulants will not be needed.

Using ~\eqref{eq:edgeworth-general} with $g_k(u)=u^{-k}$, whose second derivative is $g_k''(u)=k(k+1)u^{-(k+2)}$,
we find, for $k=2$ and $k=4$,
\begin{align}
\mathbb{E}\!\left[\frac{1}{\bar W^{\,2}}\right]
&= \mu^{-2}
  + \frac{3\sigma^2}{N}\,\mu^{-4}
  - \frac{4\kappa_3}{N^2}\,\mu^{-5}
  + \frac{15\sigma^4}{N^2}\,\mu^{-6}
  + O(N^{-3}),
\label{eq:EWinv2}\\[4pt]
\mathbb{E}\!\left[\frac{1}{\bar W^{\,4}}\right]
&= \mu^{-4} + \frac{10\sigma^2}{N} \mu^{-6}+ O\!\left(\frac{1}{N^2}\right).
\label{eq:EWinv4}
\end{align}

\subsection{Final large-$N$ expansion of $\mathrm{MSE}$}
Recalling~\eqref{eq:mseE-core-corrected}, we obtain
\begin{equation}
\mathrm{MSE}_E
\;=\;
\frac{1}{2N}\,\mathbb{E}\!\left[\frac{1}{\bar W^{\,2}}\right]
\;-\;
\frac{1}{2N^2}\,\mathbb{E}\!\left[\frac{1}{\bar W^{\,4}}\right]
+ O(N^{-3}).
\end{equation}
Substituting the expansions \eqref{eq:EWinv2}, \eqref{eq:EWinv4} yields
\begin{equation}
\mathrm{MSE}_E
\;=\;
\frac{1}{2N}\,\frac{1}{\mu^{2}}
\;+\;
\frac{1}{2N^{2}}\,\frac{3\sigma^{2}-1}{\mu^{4}}
\;+\;
O\!\left(N^{-3}\right).
\label{eq:mseE-final}
\end{equation}
Equivalently, in terms of $(r,p)$,
\begin{equation}
\mathrm{MSE}_E
\;=\;
\frac{1}{2N\,r^{2}(1-2p)^{2}}
\;+\;
\frac{6r^{2}p(1-p)+\tfrac{1}{4}}
     {N^{2}\,r^{4}(1-2p)^{4}}
\;+\;
O\!\left(N^{-3}\right).
\label{eq:mseE-final-rp}
\end{equation}

\end{document}